\documentclass[11pt]{article}

\usepackage{amsmath,amssymb}
\usepackage{slashed}
\usepackage{xcolor} 
\usepackage{graphicx}
\usepackage{dcolumn} 
\usepackage{bm} 
\usepackage{epsfig}
\usepackage{epstopdf}
\usepackage{grffile}
\usepackage{color}
\usepackage{colordvi}
\usepackage{amsmath,amssymb}
\usepackage{rotating}
\usepackage{lscape}
\usepackage{cite}
\usepackage{float}
\usepackage{hyperref}

\def\Re{{\cal R \mskip-4mu \lower.1ex \hbox{\it e}\,}}
\def\Im{{\cal I \mskip-5mu \lower.1ex \hbox{\it m}\,}}

\def\tev{\,{\ifmmode\mathrm {TeV}\else TeV\fi}}
\def\gev{\,{\ifmmode\mathrm {GeV}\else GeV\fi}}
\def\mev{\,{\ifmmode\mathrm {MeV}\else MeV\fi}}

\begin{document}

\begin{center}

\vspace*{15mm}
\vspace{1cm}
{\Large \bf Constraints on top quark flavor changing neutral currents  using diphoton events at the LHC}

\vspace{1cm}

{\bf  Sara Khatibi  and Mojtaba Mohammadi Najafabadi }

 \vspace*{0.5cm}

{\small\sl School of Particles and Accelerators, Institute for Research in Fundamental Sciences (IPM) P.O. Box 19395-5531, Tehran, Iran } \\

\vspace*{.2cm}
\end{center}

\vspace*{10mm}

%
%
\begin{abstract}\label{abstract}
In this paper we show that the diphoton mass spectrum in proton-proton collisions at the LHC is sensitive to the top quark flavor changing neutral
current in the vertices of $tu\gamma$ and $tc\gamma$. 
The diphoton mass spectrum measured by the CMS experiment at the LHC at a
center-of-mass energy of 8 TeV and an integrated luminosity of 19.5 fb$^{-1}$ is used as an example 
to set limits on these FCNC couplings. 
It is also shown that the angular distribution of the diphotons is sensitive to
anomalous $tu\gamma$ and $tc\gamma$ couplings and it is a powerful tool to  probe any value of the
branching fraction of top quark rare decay to an up-type quark plus a
photon down to the order of $10^{-4}$. We also show that the $tu\gamma$ FCNC coupling has a significant
contribution to the neutron electric dipole
moment (EDM) and the upper bound on neutron EDM can be used to constrain the $tu\gamma$ FCNC coupling. 

\end{abstract}

\vspace*{3mm}



\newpage


\section{Introduction}\label{Introduction}

The top quark with a mass of $173.34 \pm 0.75$ GeV  \cite{pdg} is the heaviest particle of the Standard Model (SM). 
With such a mass, the top quark has the largest Yukawa coupling to the Higgs boson and therefore
measurement of its properties would provide a promising way to probe the electroweak
symmetry breaking mechanism and new physics beyond the SM.
New physics can
show up either through direct production of new particles or indirectly
via higher order effects. Observing indirect evidences is important as
it provides hints to look for new physics before direct discovery.
In the Standard Model (SM), the branching fractions of top quark rare
decays $t\rightarrow q V$, with $q = u,c$ and $V=\gamma,Z,g$, are
at the order of $10^{-14}-10^{-12}$ \cite{saav1}. Such branching fractions are
extremely small and are out of the ability of the current and future
collider experiments to be measured. Within the SM, such 
Flavor Changing Neutral Current (FCNC) transitions only occur at loop level and are strongly suppressed due
to the Glashow-Iliopoulos-Maiani (GIM) mechanism \cite{gim}. On the other hand, it has been shown that
several extensions of the SM are able to relax the GIM suppression of
the top quark FCNC transitions due
to additional loop diagrams mediated by new particles. Models, 
such as supersymmetry, two Higgs doublet models, predict significant
enhancements for the FCNC top quark rare decays 
\cite{bsm0,bsm1,bsm2,bsm3,bsm4,bsm5,bsm6,bsm7,bsm8,bsm9,bsm10,bsm11,bsm12,bsm13,bsm14}. As a result, the
observation of any excess for these rare decays would be indicative
of indirect effects of new physics. Many studies on searches for the
top quark FCNC and other anomalous couplings have been already done 
\cite{ee3,ewwwe1,ee2,Khatibi:2014via,Agram:2013koa,AguilarSaavedra:2000db,Larios:2004mx,Hesari:2014eua, Khanpour:2014xla, Craig:2012vj,etesami,x1,x5,x7,x8,x9,htt,wp}.

In this paper, a direct  search for the top quark FCNC interactions in the vertex
of $tq\gamma$ is discussed. Such interactions can be described in a model-independent 
way using the effective Lagrangian approach, which has the
following form \cite{lag}:
 \begin{equation}\label{eff}
 \mathcal{L}_{\rm FCNC} = - eQ_{t} \sum_{q=u,c}  \kappa_{tq\gamma} \bar{q} (\lambda^v_{tq\gamma}+\lambda^a_{tq\gamma} \gamma_{5}) \frac{i \sigma_{\mu \nu} q^{ \nu}}{\Lambda}  t A^{ \mu} +h.c.,
 \end{equation}
where the electric charges of the electron and top quark are denoted
by $e$ and $eQ_{t}$, respectively and $q^{\nu}$ is the
four momentum of the involved photon, $\Lambda$ is the cutoff of the
effective theory, which is conventionally assumed to be equal to the top
quark mass, unless we mention.  In the FCNC Lagrangian in Eq.\ref{eff},  $\sigma_{\mu \nu} = \frac{1}{2}[\gamma_{\mu},\gamma_{\nu}]$ and
the anomalous couplings strength is denoted by $\kappa_{tq\gamma}$. Throughout this paper,  no specific chirality is assumed for the
$tq\gamma$ FCNC couplings, i.e. $\lambda^v_{tq\gamma} = 1$ and $
\lambda^a_{tq\gamma}=0$. 
Within the SM framework, the values of $\kappa_{tq\gamma}$, $q=u,c$, vanish at tree level.

The leading order (LO) partial width of the top quark FCNC decay $t\rightarrow q\gamma$,
neglecting the masses of the up and charm quarks, has the
following form \cite{width}:
\begin{eqnarray} 
\Gamma (t \rightarrow q \gamma)  =  \frac{\alpha}{2} Q^{2}_{t} m_{t} |\kappa_{tq\gamma}|^{2},
\end{eqnarray}
and the LO width of $t\rightarrow  b W^{+}$ can be written as \cite{width,mmn}:
\begin{eqnarray}
\Gamma(t \rightarrow b W^{+}) = \frac{\alpha |V_{tb}|^{2}}{16
  s_{W}^{2}}\frac{m_{t}^{3}}{m_{W}^{2}} \left(1-\frac{3m_{W}^{4}}{m_{t}^{4}}+\frac{2m_{W}^{6}}{m_{t}^{6}} \right),
\end{eqnarray}
where $\alpha$ and $V_{tb}$ are the fine structure constant and the CKM
matrix element, respectively. The sine of the Weinberg angle is
denoted by $s_{W}$ and $m_{t},m_{W}$ are the top quark and $W$ boson mass,
respectively. The branching fraction of $t\rightarrow q \gamma$ is estimated as the ratio of
$\Gamma (t \rightarrow q \gamma)$ to the width of $t\rightarrow bW^{+}$
which takes the following form \cite{width}: 
\begin{eqnarray}\label{branchingratio}
Br(t\rightarrow q \gamma) =  0.2058\times |\kappa_{tq\gamma}|^{2}.
\end{eqnarray}
To obtain the above branching fraction, we set $m_{t}$=172.5 GeV,
$\alpha=1/128.92$, $m_{W}$= 80.419 GeV and $s_{W}^{2} = 0.234$ in
$t\rightarrow q\gamma$ and  $t\rightarrow bW^{+}$ widths.

The $tu\gamma$ and $tc\gamma$ FCNC couplings have been studied in different experiments with
no observation of any excess above the SM expectation up to now.
In $p\bar{p}$ collisions at the Tevatron, the CDF experiment has set
the following upper bounds on the branching fraction at  the $95\%$
confidence level  (CL) \cite{cdf}:
\begin{eqnarray} 
Br(t \rightarrow q \gamma)  < 3.2 \times 10^{-2}~,~ \text{with}~q =
  u, c.
\end{eqnarray}
This upper bound has been obtained using the study of the top quark
decays in top quark pair production.
Further searches for the anomalous $tq\gamma$ couplings in 
electron-positron and electron-proton colliders (LEP and HERA) have provided
the following limits on the anomalous couplings at the $95\%$ CL \cite{delphi,zeus,h1}:
\begin{eqnarray}
\kappa_{tc\gamma} < 0.486~\text{(DELPHI)}~,~\kappa_{tu\gamma} <
0.174~\text{(ZEUS)}~,~\kappa_{tu\gamma} < 0.18~\text{(H1)}.
\end{eqnarray}
The ZEUS limit has been
obtained under the assumption of $m_{t} = 175$ GeV.

The most stringent bounds on
the $tq\gamma$ FCNC interactions have been obtained by the CMS experiment 
at the LHC, using proton-proton collisions
 at a center-of-mass energy of 8 TeV,  by studying
the final state of single top quark production in association with a
photon. The following upper bounds have been obtained on the anomalous
couplings and the corresponding branching fractions at the $95\%$ CL \cite{cmspas}:
\begin{eqnarray}\label{cms}
\kappa_{tu\gamma} < 0.028  ~,\text{corresponding to}~Br(t\rightarrow u\gamma) < 1.61 \times
10^{-4},  \nonumber \\ 
\kappa_{tc\gamma} < 0.094 ~,\text{corresponding to}~Br(t\rightarrow c\gamma) <   1.82 \times
10^{-3}.
\end{eqnarray}
These limits has been obtained based on 19.1 fb$^{-1}$ of integrated
luminosity of data using only the muonic decay mode of the W boson in the
top quark decay.

All the above searches are based on final states containing at
least a top quark. As the top quark
has a short lifetime, it decays immediately (before hadronization). Therefore one has to
reconstruct top quark from its decay products to be able to probe the $tq\gamma$ couplings. This needs
a careful attention to correctly select the final state objects,
i.e. top quark decay products,  and 
consider several sources of systematic uncertainties associated to each final
state object in the detector.  In this work, we propose instead to use diphoton events to probe the $tq\gamma$
FCNC couplings which have less difficulties and challenges with respect to
the events with top quarks in the final state.

The measurement of the diphoton invariant mass spectrum is one of the particular interests at
the LHC as it is sensitive to several new physics models beyond the SM
\cite{cmsdiphoton1,cmsdiphoton2, atlasdiphoton1, atlasdiphoton2}, 
being one of the most sensitive channels to the Higgs boson production at the LHC. 
On the other hand, the excellent mass resolution of the diphoton spectrum in the ATLAS
and CMS detectors at the LHC provides the possibility for precise
measurement of new signals above the SM expectation. Randall-Sundrum
model \cite{rs} and large extra dimensions \cite{add}
are of the examples of the models which affect the diphoton
differential cross sections. In this paper, we show that the presence
of the FCNC anomalous coupling 
$tq\gamma$ leads  to significant change in the diphoton mass spectrum
and the diphoton angular distribution. Using a mass spectrum
measurement by the CMS experiment \cite{cmsdiphoton1}, we obtain bounds on the anomalous
couplings $\kappa_{tq\gamma}$. In addition it is shown that the diphoton angular
distribution would be able to constrain the $tq\gamma$ FCNC couplings
strongly. 

As an  indirect way to probe the FCNC couplings, we calculate
the effect of $tu\gamma$ coupling to the neutron electric dipole
moment (EDM) and show that the neutron EDM can receive significant
contribution from the FCNC couplings.

This paper is organized as follows. In Section \ref{diphotonmass}, the
details of the calculations and methods to constrain the $tq\gamma$
FCNC couplings using the diphoton mass spectrum are
presented. Section \ref{diphotonangular} is dedicated to present the
the potential of the LHC to study the $tq\gamma$
FCNC couplings using the angular distribution of the diphoton events.
Finally, Section \ref{Conclusions} concludes the paper.  In the appendix \ref{App},
using the upper bound on the neutron EDM, an upper limit on the anomalous $tu\gamma$ is derived.

\section{ Diphoton: mass spectrum}\label{diphotonmass}

In this section, we calculate the contribution of $tq\gamma$ FCNC
couplings to diphoton production at the LHC. Then, based on the
measured diphoton mass spectrum by the CMS experiment \cite{cmsdiphoton1}, constraints on
the anomalous couplings are derived. 

\begin{figure}[htb]
\begin{center}
\vspace{1cm}
\resizebox{0.55\textwidth}{!}{\includegraphics{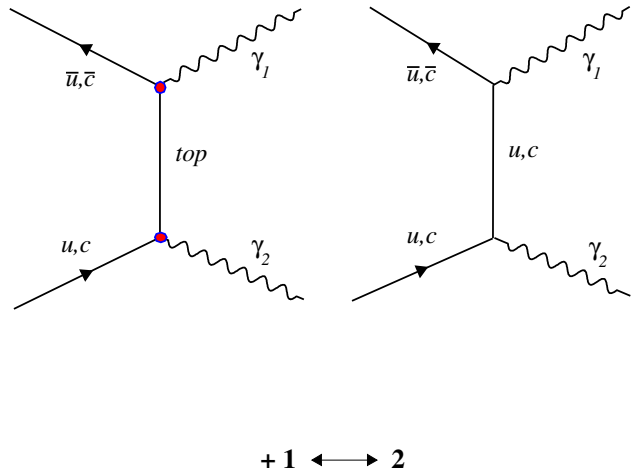}}  
\caption{ The representative Feynman diagrams of the $tq\gamma$ FCNC
  contributions to the diphoton production at the LHC. The right
  diagram  represents the lowest order SM contribution to
  diphoton production which interferes with diagrams from $tq\gamma$. }\label{fig:feynman}
\end{center}
\end{figure}

Within the SM, the LO diphoton production proceeds through
quark-antiquark annihilation.
The $tq\gamma$ FCNC couplings affect the diphoton production through
the scattering of $u,c,\bar{u}$, and $\bar{c}$ quarks which proceed through
$t$-channel as shown in Fig. \ref{fig:feynman}.

 We calculate the leading order matrix
element of diphoton production analytically for the Feynman diagrams shown in
Fig.\ref{fig:feynman}. After averaging over the color and spin indices of the
initial state partons and summing over the polarizations of the final
state photons, the amplitude takes the following form:
\begin{eqnarray}
\overline{|\mathcal{M}|}^2 \propto \frac{2 e^{4}u}{t}+\frac{ Q_{t}^{2}e^{4}\kappa^{2}t (u-4
  t)}{m^2_{t}(t - m^2_{t})} +\frac{2Q_{t}^{4}  e^{4} \kappa^{4}t^2 (m_{t}^2 s+t u)}{m^4_{t}(t - m^2_{t})^{2}} ,
\end{eqnarray}
where for simplicity, we have assumed
$\kappa_{tu\gamma}=\kappa_{tc\gamma}=\kappa$ and $s,t,u$ are the
Mandelstam variables which can be written in terms of the scattering
angle $\theta^{*}$ in the center-of-mass frame as: $t =
-\frac{s}{2}(1-\cos\theta^{*})$ and $u= -\frac{s}{2}(1+\cos\theta^{*})$. The first term in the
above expression is the leading order amplitude describing the SM
diphoton production, the
second term is the interference between the SM and FCNC diagram, and
the last term is the contribution of pure $tq\gamma$ FCNC diphoton
production.  The interference term (SM+FCNC) is found to be constructive and the contribution of the third term, 
which is purely coming from FCNC, is smaller than the
the interference term by a factor of $\approx 10^{-3}$. 
One of the characteristics of the LO SM is the
enhancement of diphoton production at small angles as the production
proceed through a $t$-channel virtual exchange.

In order to perform the signal simulation, the $tq\gamma$ effective
Lagrangian, Eq.\ref{eff}, is implemented into the
{\sc FeynRules} package \cite{feynrule} and then the model is exported to a UFO
module \cite{ufo}  which is linked to {\sc MadGraph 5} \cite{mg,mg2}. Events are generated, describing
the diphoton production at the LHC with the center-of-mass energy of
$\sqrt{s} = 8$ TeV.  The LO parton distribution functions (PDFs)
of CTEQ6L1 \cite{cteq} are used as the input for the calculations and events
generation. The renormalization and factorization scales are chosen to be equal and 
set to the default dynamic scales of the  {\sc MadGraph}  generator.
{\sc Pythia 8} \cite{pythia} is used for parton showering and
hadronization of the parton-level events.
Finally,  the detector-level effects are emulated by
{\sc Delphes-3.3.2} package \cite{delphes}. It includes
a reasonable modeling of the CMS detector performances as
described in \cite{cmsdetector}.

In \cite{cmsdiphoton1}, the CMS collaboration has performed a
search for diphoton resonances in high mass in proton-proton
collisions at the center-of-mass energy of 8 TeV using an integrated
luminosity of 19.5 fb$^{-1}$ of data. The analysis searches for resonant diphoton production via 
gravitons in the Randall-Sundrum scenario with a
warped extra dimension. According to the calculations presented above, the
$tq\gamma$ FCNC couplings affect the production of diphotons at the
LHC. In this work, we follow the quite similar strategy to the CMS
collaboration and use their result to probe the $tq\gamma$ FCNC
anomalous couplings.

In the CMS experiment analysis, two isolated photons with transverse energy ($E_{T}$) greater
than 80 GeV within the pseudorapidity range of $|\eta_{\gamma}| <
1.4442$, and with  a diphoton-system invariant mass greater than 300 GeV
are selected. In this region of the pseudorapidity, an excellent
resolution for the photon energy is experimentally achieved. For the photons with
$E_{T} \sim 60$ GeV and $|\eta_{\gamma}| < 1.4442$, the energy resolution
varies between $1\%-3\%$ \cite{photon}.
The used cuts for isolation and identification of the
photons by the CMS collaboration lead to an efficiency of $86\%$ for
the photons with $E_{T} > 80$ GeV and  $|\eta_{\gamma}|
<1.4442$. Small changes are seen in this efficiency when the $E_{T}$ and
$\eta$ of the photons change. In the current work, quite similar
selection is employed for the analysis \cite{cmsdiphoton1,cmsdiphoton2}.

The background to the diphoton final state originates from SM diphoton
production, $\gamma$+jet, and from dijet productions where one or two
jets are misidentified as photons in the detector for the latter two
background processes. Table
\ref{cmstable} shows the number of observed events in data and the
background prediction for several ranges of the diphoton mass
spectrum \cite{cmsdiphoton1}. The uncertainties presented in the Table \ref{cmstable}
include both the statistical and systematic sources.
The data and SM background expectation
are found to be in agreement, considering the uncertainties on the
predicted background and no significant excess over the SM background
is found. 

\begin{figure}[htb]
	\begin{center}
		\vspace{1cm}
		\resizebox{0.70\textwidth}{!}{\includegraphics{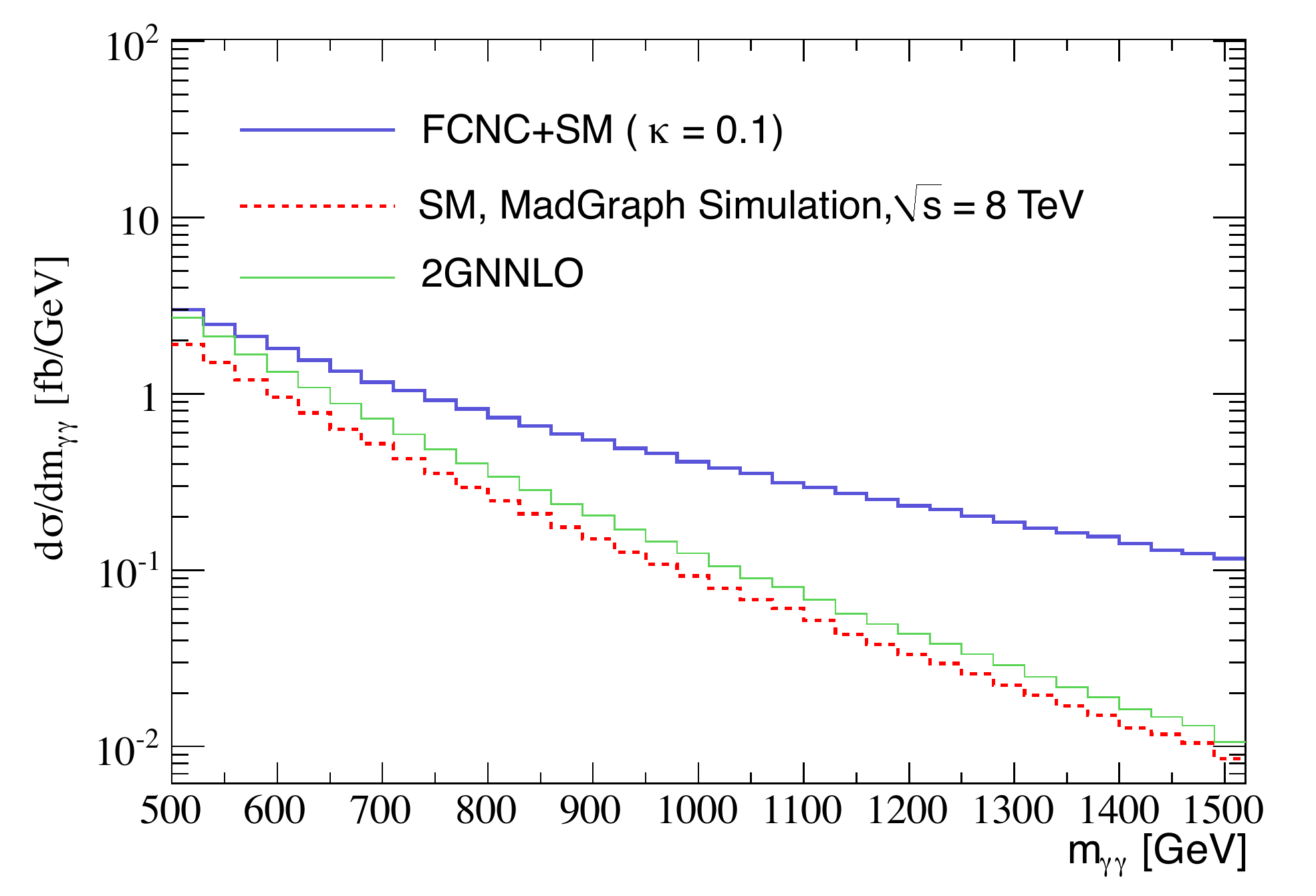}} 
		\caption{Diphoton invariant mass distribution for SM
                  and SM+FCNC with $\kappa = 0.1$ obtained from
                  LO MadGraph simulation at the
                  center-of-mass energy of 8 TeV. The SM prediction at
                  NNLO obtained from {\sc 2gNNLO} is also depicted for comparison. 
                      }
		\label{massgg}
	\end{center}
\end{figure}

The values reported in Table \ref{cmstable} are used to probe $tq\gamma$ anomalous
couplings. As the measurement is compatible with the SM prediction, we
set upper limit on the diphoton production cross section in the
presence of anomalous couplings.
Figure \ref{massgg} shows the diphoton mass distribution at LO for the SM and
SM+FCNC signal assuming $\kappa_{tu\gamma} = \kappa_{tc\gamma}=\kappa
= 0.1$ obtained from the {\sc MadGraph} simulation. The diphoton mass
distribution at NNLO estimated based on the Monte Carlo program {\sc
  2gNNLO} \cite{2g1,2g2} is also shown in Fig.\ref{massgg} for
comparison.
{\sc 2gNNLO} program calculates the production 
cross section of diphoton  in hadron collisions to the accuracy of  next-to-next-to-leading-order.
As depicted, the presence of $tq\gamma$ FCNC couplings lead
to increase the diphoton cross section in the high invariant mass
region. According to Table \ref{cmstable}, the total number of
observed data events above $m_{\gamma\gamma} > 500$ GeV is 333 events
with the SM background prediction of $375.8 \pm 29.9$ \cite{cmsdiphoton1}, where the SM diphoton production has been 
estimated based on the Monte Carlo program {\sc 2gNNLO}. 
Assuming  $\kappa_{tu\gamma} = \kappa_{tc\gamma}=\kappa
= 0.15$, $99.4 \pm 8.5$ FCNC events are expected in this region for an
integrated luminosity of 19.5
fb$^{-1}$ of data.  The uncertainty on the number of FCNC events includes the contributions coming from 
the choice PDFs, variations of renormalization and factorization scales, and the statistical uncertainty.
The PDF uncertainty is obtained according to the PDF4LHC recommendation \cite{pdf4lhc1,pdf4lhc2} using PDF sets 
CTEQ6L1 \cite{cteq}, NNPDF 3.0 \cite{nnpdf}, and MSTW 2008 \cite{mstw}. The uncertainty originating from the 
variations of the scales has been estimated by varying the renormalization and factorization scales simultaneously by factors of 0.5 and 2.

\begin{table}[htbp]
\centering
\caption{Number of observed events in data and the SM background
  prediction in different ranges of diphoton invariant mass with 19.5
  fb$^{-1}$ of data \cite{cmsdiphoton1}. }
\label{cmstable}
\begin{tabular}{|c|c|c|c|}
\hline
$   m_{\gamma\gamma}  $ [GeV] & Data & Total expected (SM) \\  \hline
500-750          & 265 & 310.8  $\pm 29.9$     \\   \hline
750-1000        & 46   & 48.6  $\pm 5.4$       \\   \hline
1000-1250      & 16   & 11.4  $\pm 1.5$    \\ \hline
1250-1500      &  3    & 3.3    $\pm 0.5$   \\  \hline
1500-1750      &  2    & 1.1     $\pm 0.2$   \\  \hline
1750-$\infty$ &  1    &  0.6    $\pm 0.1$   \\ \hline
\end{tabular}
\end{table}

\begin{figure}[htb]
	\begin{center}
		\vspace{1cm}
		\resizebox{0.70\textwidth}{!}{\includegraphics{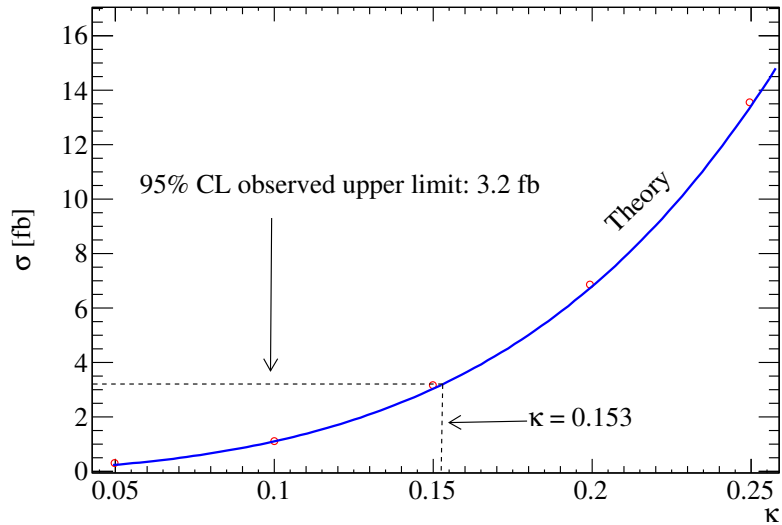}}  
		\caption{ Parameterization of the signal cross section
                  versus the anomalous FCNC coupling $\kappa$ and the
                  observed $95\%$ CL upper limit on the cross section. }
		\label{sigmalm}
	\end{center}
\end{figure}

We proceed to set an upper limit on the diphoton cross section in the
presence of FCNC couplings. We compare the number of
observed events in data and the expected events from SM in the region
$m_{\gamma\gamma} > 500$ GeV. The limit at the $95\%$ CL is set on the
quantity $\sigma_{s} = (\sigma_{\rm Total} - \sigma_{\rm SM})\times\epsilon_{A}$, where the
whole diphoton production cross section (SM and FCNC signal) is
denoted by $\sigma_{\rm Total}$ and $\sigma_{\rm SM}$ is the  SM  diphoton
cross section. The FCNC signal acceptance is taken into account by the  $\epsilon_{A}$ term.
The CL$_{s}$ technique \cite{cls} is used to calculate the upper limit on the cross
section. An efficiency of $77.45 \%$ with an
uncertainty of $10\%$ is found for the FCNC signal.  The observed and
expected $95\%$ CL upper limit on $\sigma_{s}$ are found to be  3.2 fb
and 5.0 fb.  The observed limit at the $95\%$ CL and the FCNC
signal cross section, $\sigma_{s}$ are shown in Fig.
\ref{sigmalm}.

The  $95\%$ CL upper limit on the
anomalous coupling parameter $\kappa$ is the intersection of the
observed limit on cross section with the theoretical cross section
curve. The upper limit on $\sigma_{s}$ (3.2 fb) is corresponding to
the upper limit of 0.153 on the anomalous coupling $\kappa$. This
limit can be expressed to the upper limit on the branching fraction
using Eq.\ref{branchingratio}:
\begin{eqnarray}
Br(t\rightarrow q\gamma) < 4.81 \times 10^{-3}, ~\text{with} ~q = u,c.
\end{eqnarray}
The value obtained is comparable to the most stringent limits which has been
obtained from the anomalous single top quark production in association
with a photon by the CMS experiment (Eq.\ref{cms}) \cite{cmspas}. 
This provides a motivation for using this channel as a complementary technique 
for studying the $tq\gamma$ FCNC interactions at the LHC experiments. A combination of
this result with the results of other channels can lead to an improvement of the best limit.

\section{Angular distribution of the diphoton system}\label{diphotonangular}

In this section, we propose and use a diphoton angular variable to
probe the $tq\gamma$ anomalous couplings. In the SM, as the diphoton
production proceeds through a $t$-channel exchange, the angular distribution
peaks at  $\cos\theta^{*} = 1$, where $\theta^{*}$ is the 
scattering angle in the center-of-mass frame of two partons.
The  scattering  angle  between  two  photons
can also be expressed by the variable $\chi =
e^{|\eta_{\gamma_{1}}-\eta_{\gamma_{2}}|}$. This variable has been used widely
in searches for new physics such as searches for contact interactions,
large extra dimensions, and excited quarks in dijet events in the
Tevatron and LHC experiments \cite{dijet1,dijet2,dijet3,dijet4}.
It has been found that new phenomena affect this angular variable
considerably and consequently is used to probe beyond SM.

\begin{figure}[htb]
	\begin{center}
		\vspace{1cm}
		\resizebox{0.60\textwidth}{!}{\includegraphics{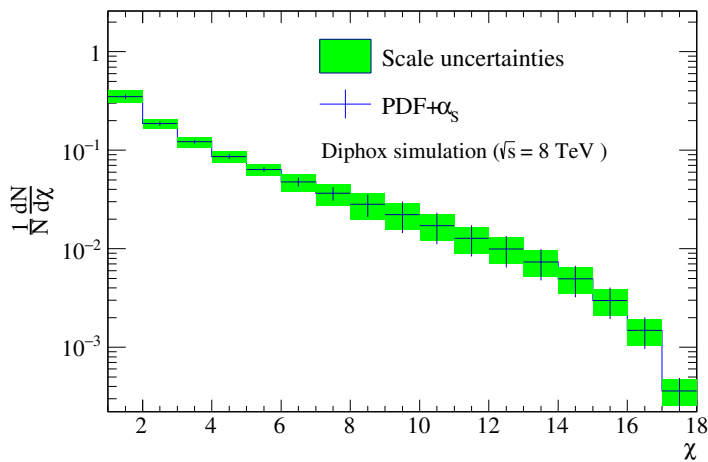}}  
		\caption{Distribution of the angular variable  $\chi$ for the SM diphoton events
                               with an invariant mass above 500 GeV and $|\eta| < 1.442$.  The error bars
represent the systematic uncertainties  coming from PDFs and strong coupling constant $\alpha_{S}$.
The shaded bars show the uncertainty from the theoretical scales in each bin of the angular distribution.}
		\label{ratioSM}
	\end{center}
\end{figure}

In order to produce the SM diphoton events, including the QCD next-to-leading order corrections and 
the contributions from the fragmentation processes,  the {\sc Diphox} (v 1.3.3) program \cite{diphox} is used. 
The CT10 PDF set  \cite{ct10} is used as the input of the parton distribution functions and all the scales are set to 
$m_{\gamma\gamma}$. Figure \ref{ratioSM}
shows the distribution of the angular variable  $\chi$ for the SM diphoton events
with an invariant mass above 500 GeV and $|\eta| < 1.442$.  The error bars
represent the systematic uncertainties  coming from the parton distribution functions and strong coupling constant $\alpha_{S}$.
The shaded bars show the uncertainty from the theoretical scales in each bin of the angular distribution.
The uncertainties arising from the scales variations  are calculated by varying the factorization, renormalization, and fragmentation scales simultaneously
by factors of 0.5 and 2. In each bin of the $\chi$ angular distribution, the uncertainty is calculated as the maximum difference between the new angular distributions and the 
distribution with the reference inputs.
 The uncertainty coming from limited knowledge on
the choice of PDF is obtained using PDF4LHC recommendation \cite{pdf4lhc1,pdf4lhc2}.
The  PDF sets CT10 \cite{ct10}, MSTW08 \cite{mstw}, and NNPDF 3.0 \cite{nnpdf} are used to estimate the uncertainty on 
from PDFs and strong coupling constant $\alpha_{S}$ is varied by 0.012 similar to the prescription adopted in \cite{cmsgamma}.

\begin{figure}[htb]
	\begin{center} 
\includegraphics[width=7cm, height=5cm]{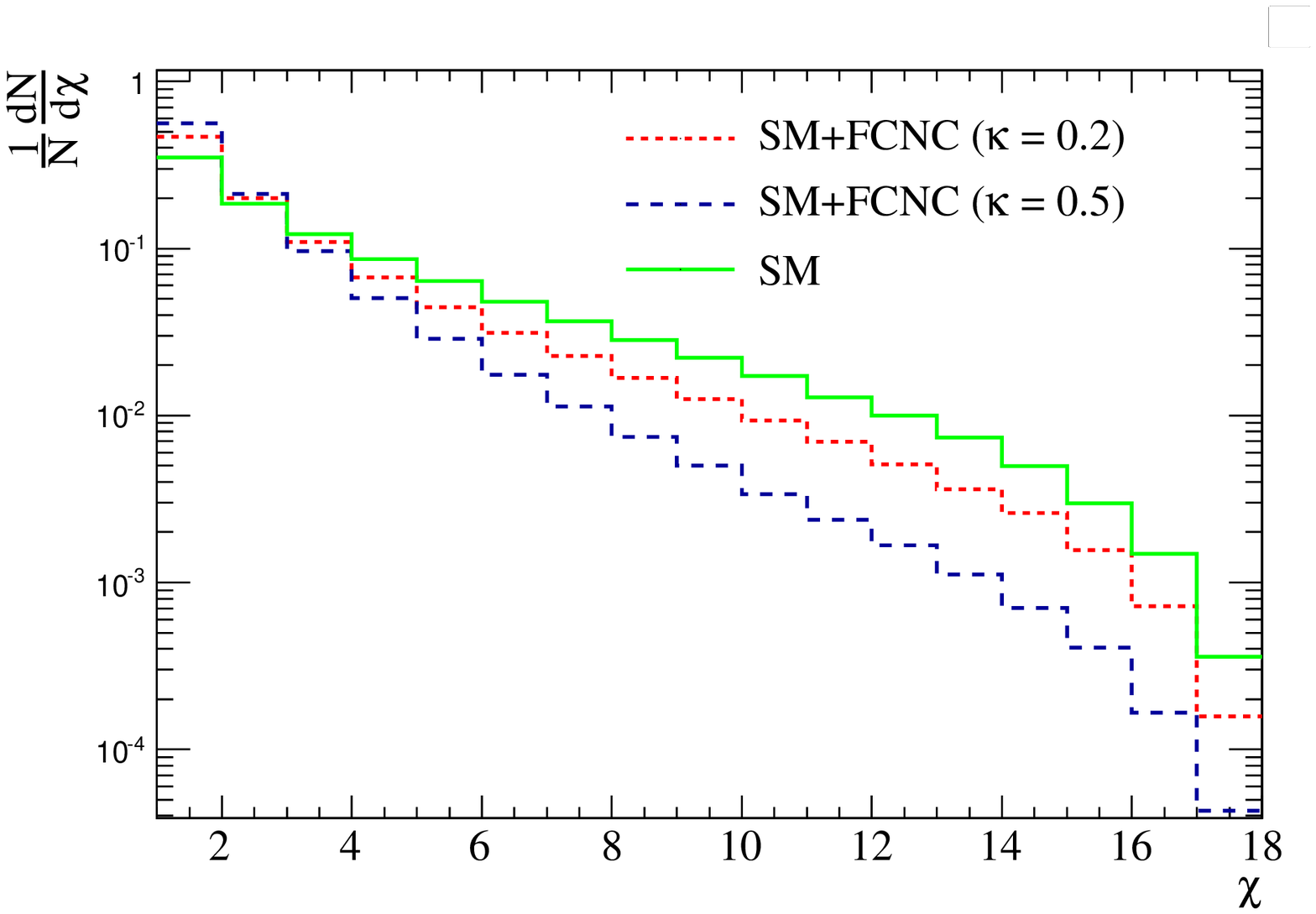}
\includegraphics[width=7cm, height=5cm]{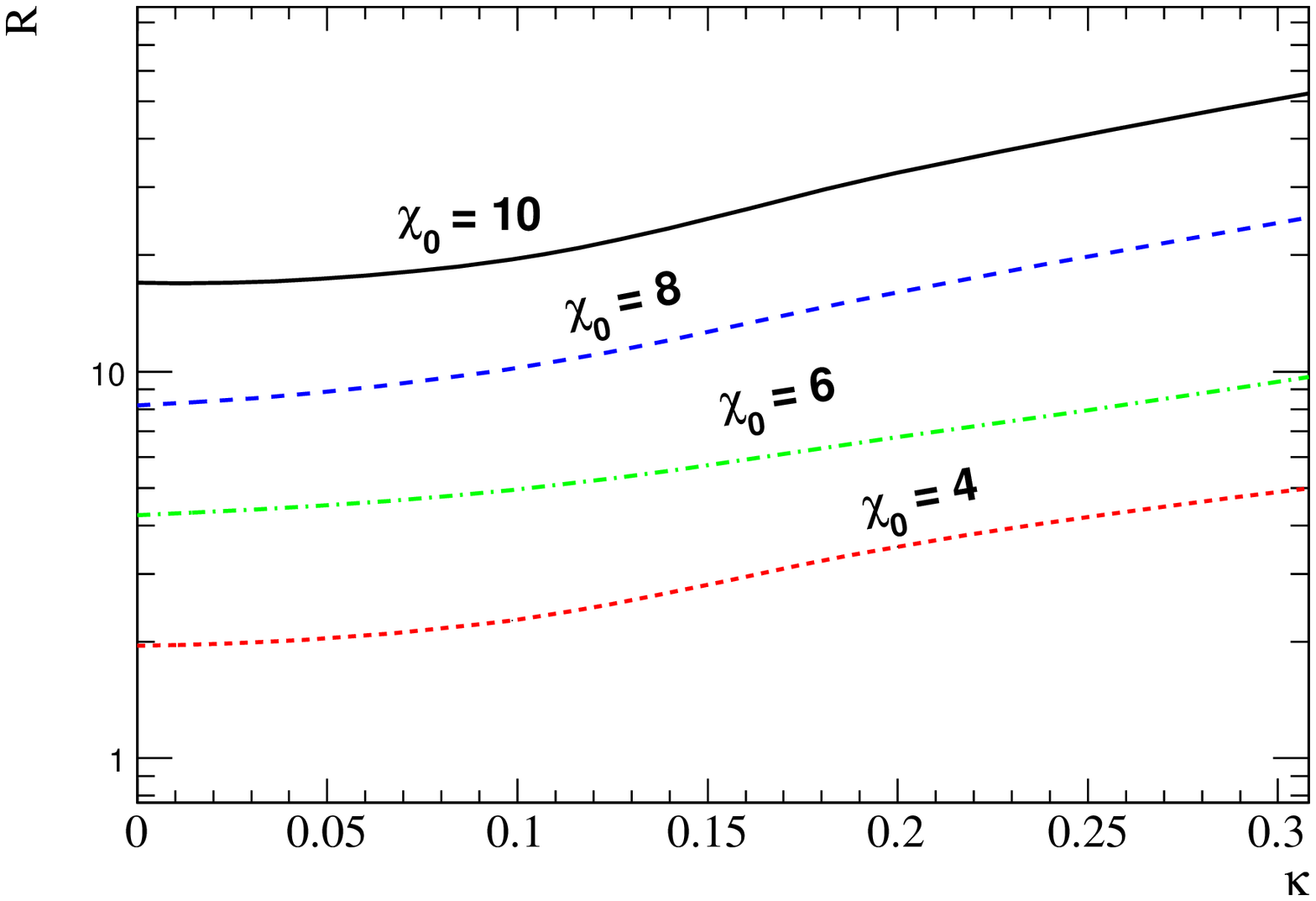}
		\caption{Left: Normalized distributions  of diphoton  angular
                  variable $\chi$ for the SM and SM+FCNC with
                  anomalous couplings $\kappa$ = 0.2,0.5. Right:
                  $\mathcal{R}$ versus the anomalous coupling
                  $\kappa$ for various choices of $\chi_{0}$.}
		\label{chi1}
	\end{center}
\end{figure}

The left plot in Fig. \ref{chi1} shows the distributions of
$\chi=e^{|\eta_{\gamma_{1}}-\eta_{\gamma_{2}}|}$ as a function of the
anomalous coupling $\kappa$. The  distribution  is  normalized to
unity since the sensitivity to FCNC couplings affects the  angular
distribution  rather  than  normalization.  This figure depicts the
predicted SM distribution as well as the SM+FCNC with $\kappa =
0.2$ and 0.5. These distributions are after all selection cuts
described previously requiring in addition that  $m_{\gamma\gamma} > 500$ GeV. As seen, the presence of the anomalous FCNC
couplings of $tq\gamma$ changes the shape of the angular variable
$\chi$. Increasing the value of the anomalous coupling $\kappa$ causes
more events to be concentrated at small values of $\chi$. It is
notable that due to the detector acceptance cut applied on the photon
pseudorapidity, $\chi $ varies from 0 to $e^{2\times 1.442} = 17.96$.

\begin{figure}[htb]
	\begin{center}
		\vspace{1cm}
		\resizebox{0.60\textwidth}{!}{\includegraphics{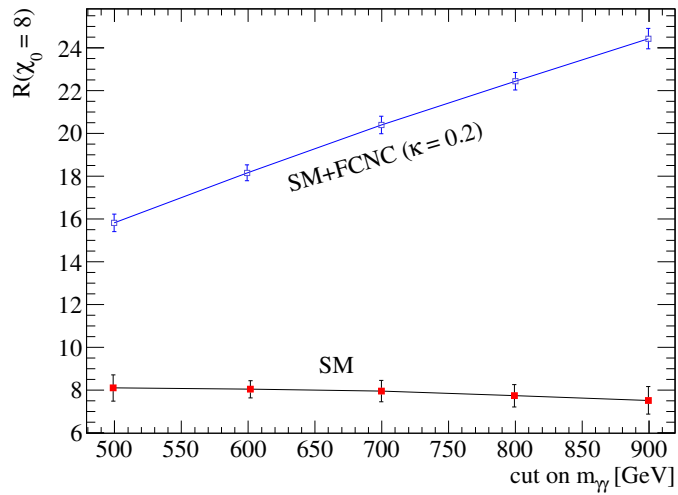}}  
		\caption{The behavior of $\mathcal{R}$ in terms of
                  the cut on diphoton mass for the SM and SM+FCNC at
                  $\chi_{0} = 8$ and with the choice of $\kappa = 0.2$.  }
		\label{ratioaa}
	\end{center}
\end{figure}

In order to quantify the difference in the shape of $\chi$ for SM and FCNC, a ratio is
defined as:
\begin{eqnarray} \label{ratioc}
\mathcal{R}_{\chi_{0}} (\kappa)=\frac{\int_{0}^{\chi_{0}}\frac{1}{N}\frac{dN}{d\chi}}{\int_{\chi_{0}}^{\infty}\frac{1}{N}\frac{dN}{d\chi}},
\end{eqnarray}
where $\chi_{0}$ is an arbitrary cut which is chosen in such a way
that the best sensitivity to the FCNC couplings is achieved.  The right plot in
Fig. \ref{chi1} shows the behavior of $\mathcal{R}$ versus the
anomalous coupling $\kappa$ for different choices of $\chi_{0}$.
As the normalized
distribution of $\chi$ depends on the cut on the diphoton invariant
mass, the value of $\mathcal{R}$ varies with the cut on
$m_{\gamma\gamma}$.  Figure \ref{ratioaa} shows the behavior of
$\mathcal{R}$ for the SM and SM+FCNC 
for the 
$\chi_{0} = 8$ choice and different cuts on the minimum $m_{\gamma\gamma}$.
  The uncertainties in this plot includes both the statistical and theoretical uncertainties for the SM 
and only the statistical uncertainty for the SM+FCNC.
The cut on $m_{\gamma\gamma}$ can be chosen to optimize the expected sensitivity
to $\kappa$.

We define the statistical significance of the observable $\mathcal{R}$ as:
\begin{eqnarray}
\mathcal{S}_{\chi_{0}} (\kappa) = \frac{\mathcal{R}_{\chi_{0}}^{\rm FCNC+SM}(\kappa)-\mathcal{R}_{\chi_{0}}^{\rm SM}}{\Delta \mathcal{R}_{\chi_{0}}^{SM}},
\end{eqnarray}
where $\mathcal{R}_{\chi_{0}}^{\rm SM}$ and
$\mathcal{R}_{\chi_{0}}^{\rm FCNC+SM}$ are the values of the ratio defined
in Eq.\ref{ratioc} with a choice of $\chi_{0}$ for the SM and for the case of the presence of
FCNC.  The uncertainty on $\mathcal{R}_{\chi_{0}}^{\rm SM}$ is denoted
by $\Delta \mathcal{R}_{\chi_{0}}^{SM}$. Considering the theoretical
and statistical uncertainties in the region of $m_{\gamma\gamma} > 500$,  the value
of $\chi_{0} =8$ is found to provide the best sensitivity. 
The upper bounds at the $68\%$ CL and at the $95\%$ CL on the FCNC anomalous couplings including only statistical uncertainties are found to be:
\begin{eqnarray}
68\%~\text{CL}:~ \kappa < 2.75\times 10^{-2}~\text{corresponding to}~Br(t\rightarrow
q\gamma) < 1.56\times 10^{-4},    \nonumber \\
95\%~\text{CL}:~ \kappa < 3.91\times 10^{-2}~\text{corresponding to}~Br(t\rightarrow
q\gamma) < 3.15\times 10^{-4},
\end{eqnarray}
and the $68\%$ and $95\%$ CL limits after including both the statistical and systematic 
uncertainties are:
\begin{eqnarray}
68\%~\text{CL}:~ \kappa < 4.46\times 10^{-2}~\text{corresponding to}~Br(t\rightarrow
q\gamma) < 4.10\times 10^{-4},    \nonumber \\
95\%~\text{CL}:~ \kappa < 6.26\times 10^{-2}~\text{corresponding to}~Br(t\rightarrow
q\gamma) < 8.06\times 10^{-4}.
\end{eqnarray}

From these results it can be concluded that the angular variable $\chi$ is able to provide additional
sensitivity to $tq\gamma$ anomalous coupling with respect to the diphoton
mass spectrum.  
Further optimization on both $m_{\gamma\gamma}$ and $\chi_{0}$ is expected to improve 
the sensitivity to possible $tq\gamma$ contribution to diphoton production at the LHC.

%
%
\section{Summary and conclusions}\label{Conclusions}
Rare top quark decays through flavor changing neutral currents in the
vertices of $tq\gamma$, $tqZ$, and $tqg$ are particularly interesting
as they are significantly sensitive to many extensions of the SM.
The SM predictions for the branching ratios of these rare decay modes are expected to be unobservable at the LHC
( $ < 10^{-12}$) while new physics models are able to enhance the
branching fractions by several order of magnitudes. As a consequence, any
observation of such processes would indicate new physics beyond the
SM. In this paper, we propose a new indirect way to search for the $tu\gamma$ and
$tc\gamma$ FCNC interactions. So far, these couplings have been directly
studied by CDF, DELPHI, H1, ZEUS, and CMS experiments at
colliders with at least a top quark in the final state of the
collisions. In this work, we propose to use the diphoton
invariant mass and angular differential distributions to probe $tq\gamma$
FCNC couplings. Using a measured mass spectrum of diphoton
at the LHC with the CMS experiment, an upper limit of $4.81 \times
10^{-3}$ is set on the branching fraction of $t\rightarrow
q\gamma$. Furthermore, we show that the angular variable
$\chi=e^{|\eta_{\gamma_{1}}-\eta_{\gamma_{2}|}}$ would allow us to
probe this branching fraction down to $8.2\times 10^{-4}$. These
limits have been obtained based on the LO prediction of the $tq\gamma$ FCNC contribution in diphoton productions at the LHC and are
comparable with the ones recently obtained from the search for anomalous single top
events by the CMS experiment. 

\appendix
\section{Electric dipole moment analysis} \label{App}

In this section, we obtain upper limit on the $tu\gamma$ FCNC coupling using the present upper bound on the neutron electric dipole moment.
Such an approach has been used in constraining the $W$ boson electric
dipole moment \cite{wedm}, top-Higgs non-standard interactions \cite{htt},
and probing heavy charged gauge boson mass and couplings \cite{wp}. 

We calculate the contribution of the $tu\gamma$
coupling to the neutron EDM using the effective interaction for the
$q\bar{q}\gamma$ vertex. The most general effective vertex
describing the interaction of a photon with two on-shell quarks can be
written as \cite{lag}:
\begin{eqnarray}\label{edmlag}
\Gamma_{\mu}(q^{2})= -ie
\left(\gamma_{\mu} F_{1v}(q^{2})   +
  \frac{\sigma_{\mu\nu}}{2m_{q}}q^{\nu}[iF_{2v}(q^{2})+F_{2a}(q^{2})\gamma_{5}]
\right),
\end{eqnarray}
where $q$ is the four-momentum of
the off-shell photon. The functions $F_{1v}(q^{2})$ and
$F_{2v,2a}(q^{2})$ are called form factors which in the low energy
limit $q^{2} \rightarrow 0$, they are physical parameters and be
related to the static physical quantities according to the following
relations:
\begin{eqnarray}
F_{1v}(0) = Q_{q}~,~F_{2v}(0) = a_{q}~,~F_{2a} = d_{q}\frac{2m_{q}}{e},
\end{eqnarray}

where $Q_{q}$ is the electric charge of a quark $q$, $a_{q}$ and
$q_{q}$ are the magnetic dipole moment and electric dipole moment of a quark. 
The electric dipole moment term violates the P and CP
invariance. Within the SM at tree level, $d_{q}$ and $a_{q}$ are zero
and however non-zero values for $d_{q}$ and $a_{q}$ arise from higher
order corrections. The SM prediction for the electric dipole moments
of the quarks are extremely small and expected to be smaller
$10^{-30}$ e.cm \cite{edm1,edm2,edm3, Pospelov:2005pr}.

\begin{figure}[htb]
	\begin{center}
		\vspace{1cm}
		\resizebox{0.50\textwidth}{!}{\includegraphics{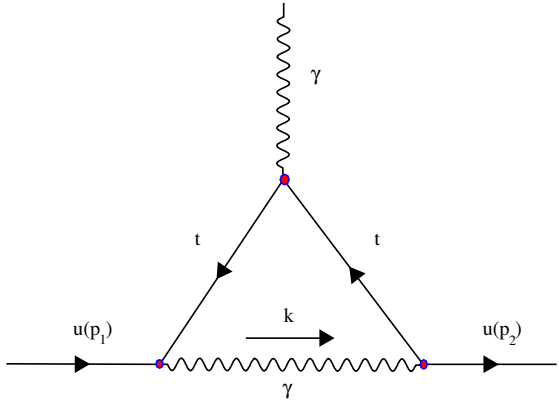}}  
		\caption{Feynman diagram contributing to the on shell $u\bar{u}\gamma$
			vertex originating from the $tuH$ interaction.}\label{loop1}
	\end{center}
\end{figure}

Using the interactions described by Eq.\ref{edmlag} at low energy, the $tu\gamma$
FCNC coupling introduced by Eq.\ref{eff}, the induced CP violating
amplitude coming from the $tu\gamma$ FCNC interaction in
Fig.~\ref{loop1} can be expressed as:

\begin{footnotesize}
\begin{eqnarray}
\Gamma_{\mu}= \bar{u}(p_{2}) [\int \frac{d^4 k}{(2 \pi)^4} (\frac{Q_{t}e \kappa_{tu\gamma}}{\Lambda} \sigma_{\beta \beta'} k^{\beta'})\frac{\imath (k\!\!\!/-p\!\!\!/_{2}+m_{t})}{(k-p_{2})^2-m^{2}_{t}} (-\imath d_{t} \gamma_{5} \sigma_{\mu \nu} q^{\nu}) \frac{\imath (k\!\!\!/-p\!\!\!/_{1}+m_{t})}{(k-p_{1})^2-m^{2}_{t}}(\frac{Q_{t}e \kappa_{tu\gamma}}{\Lambda} \sigma_{\alpha \alpha'} k^{\alpha'})\frac{-\imath g^{\alpha \beta}}{k^2} ] u(p_{1}).
\end{eqnarray}
\end{footnotesize}

After employing Dirac equation and Gordon identity, the above expression can be
simplified to find the up quark EDM arising from $tu\gamma$ FCNC coupling, which is the coefficient of
$\sigma_{\mu\nu}q^{\nu}\gamma_{5}$. 
This integral on $k$ is divergent, therefore a mass scale of $\Lambda_{cut}$
is introduced as an ultraviolet cutoff scale. 
After performing some algebraic
manipulations and integration over $k$, the amplitude is found to be:
\begin{footnotesize}
\begin{eqnarray}
\Gamma_{\mu} &=&  \frac{d_{t} (\frac{Q_{t}e \kappa_{tu\gamma}}{\Lambda})^2 m^2_{t}}{(4 \pi)^2}   [\bar{u}(p_{2}) \imath\gamma_{5} \sigma_{\mu \nu} q^{\nu} u(p_{1})] \int 2 dx dy  \nonumber \\
&\times & \lbrace \frac{- \Lambda_{\rm cut} ^2}{m^2_{t} }+ \ln \frac{\Lambda_{\rm cut} ^2}{\Delta}[6 x_{u} (x+y)(x+y-1)+3(x+y)+x_{u}-1]+\frac{x_{u}(x+y)}{2(1+x_{u}(x+y-1))} \rbrace,
\end{eqnarray}
\end{footnotesize}
where $d_{t}$ is the top quark EDM and 
\begin{eqnarray}
\Delta = m^2_{t}(x+y)[1+ x_{u}(x+y-1)]~,
~x_{u} = \frac{m^2_{u}}{m^2_{t}}. \nonumber
\end{eqnarray}
As $x_{u}\sim 10^{-5}$, we take the limit of the amplitude for the
case of $x_{u} \rightarrow 0$. We find the following form for the amplitude:

\begin{eqnarray}
\Gamma_{\mu} &=&  \frac{d_{t} e^2 Q_{t}^{2}\kappa^2_{tu\gamma}  m^2_{t}}{(4 \pi)^2 \Lambda ^2}   [\bar{u}(p_{2}) \imath\gamma_{5} \sigma_{\mu \nu} q^{\nu} u(p_{1})]  \lbrace \frac{1}{6} - \frac{ \Lambda _{\rm cut} ^2}{m^2_{t} }+ \ln \frac{\Lambda_{\rm cut} ^2}{m^2_{t} } \rbrace.
\end{eqnarray}

Employing the effective coupling for $u\bar{u}\gamma$ coupling, we find
the $tu\gamma$ contribution to the up quark EDM:
\begin{equation}
d_{u}= \frac{d_{t} e^2Q_{t}^{2} \kappa^2_{tu\gamma}  }{(4 \pi)^2 } \frac{m^2_{t}}{\Lambda ^2} \lbrace \frac{1}{6} - \frac{ \Lambda _{\rm cut} ^2}{m^2_{t} }+ \ln \frac{\Lambda_{\rm cut} ^2}{m^2_{t} } \rbrace.
\end{equation}
As seen, there are quadratic and logarithmic divergences to the up
quark EDM from $tu\gamma$ FCNC. However, there is a factor
$\Lambda^{2}$ in the denominator which comes form the $tu\gamma$
effective coupling and is the scale at which new physics effects is
expected to appear. It is natural to assume that the cutoff scale of
the loop divergences ($\Lambda_{\rm cut}$) is equal to $\Lambda$ which
is the scale at which new physics effects are expected to show
up, i.e. $\Lambda_{\rm cut} = \Lambda$.

Using the non-relativistic SU(6) wave functions, the neutron EDM can
be related to the up and down quark EDMs. 
The neutron EDM in terms of EDMs of quarks is written as~\cite{Pospelov:2005pr}:
\begin{eqnarray}
d_{n} = \eta (\frac{4}{3}d_{d} - \frac{1}{3}d_{u}) \,,
\end{eqnarray}
where the up and down quarks EDMs are denoted by $d_{u}$ and $d_{d}$
and $\eta$ describes the QCD higher order
corrections and is equal to 0.61.
The current experimental bound on the neutron EDM is $d_{n} < 2.9\times 10^{-26}$ e.cm. \cite{Altarev:1992cf,Altarev:1996xs}.
The measured upper limit on the top quark EDM has been found to be $d_{t} < 10^{-16}$
e.cm \cite{topedm}.  For example, by setting $d_{t}  = 10^{-16}, \Lambda = m_{t}$  the upper bound of 0.0026 on $\kappa_{tu\gamma}$ is obtained.
This is corresponding to an upper limit of  the order of $10^{-6}$ on 
$Br(t\rightarrow u\gamma) $.

%
%
%

%
%

\end{document}